\documentclass[%
 reprint,pra,
superscriptaddress,
 amsmath,amssymb,aps,
]{revtex4-2}

\usepackage{graphicx}% Include figure files
\usepackage{dcolumn}% Align table columns on decimal point
\usepackage{bm}% bold math
\usepackage{hyperref}% add hypertext capabilities
\usepackage[utf8]{inputenc}
\newcommand{\sub}[1]{\ensuremath{_{\!\scriptscriptstyle #1}}}
\newcommand{\subt}[1]{\ensuremath{_{\scriptscriptstyle\text{#1}}}}
\renewcommand{\sup}[1]{\ensuremath{^{\scriptscriptstyle{#1}}}}
\newcommand{\supt}[1]{\ensuremath{^{\scriptscriptstyle\text{#1}}}}
\newcommand{\oper}[1]{{\hat{#1}}}
\newcommand{\operd}[1]{{\hat{#1}}^{\dagger}}

\newcommand{\abs}[1]{\ensuremath{\left|#1\right|}}

\renewcommand{\vec}[1]{{\boldsymbol{\mathrm{#1}}}}
\newcommand{\vecp}[1]{\boldsymbol{\mathrm{#1}}^{\prime}}
\newcommand{\tp}{\ensuremath{t^{\prime}}}

\newcommand{\uvec}[1]{\ensuremath{\hat{\vec{#1}}}}
\newcommand{\uvecp}[1]{\ensuremath{\hat{\vec{#1}}^{\prime}}}

\newcommand{\omegatil}{\ensuremath{{\tilde{\omega}}}}

\newcommand{\dyad}[1]{\bar{\bar{#1}}}
\newcommand{\ket}[1]{\ensuremath{\left|{#1}\right\rangle}}
\newcommand{\bra}[1]{\ensuremath{\left\langle{#1}\right|}}

\newcommand{\proj}[1]{\ensuremath{\left|{#1}\right\rangle\left\langle{#1}\right|}}

\newcommand{\curl}[1]{\vec{\nabla}\!\times\!#1}

\newcommand{\half}{\frac{1}{2}}

\newcommand{\deriv}[2]{\frac{d #1}{d #2}}

\newcommand{\derivp}[2]{\frac{\partial #1}{\partial #2}}
\newcommand{\dderivp}[2]{\frac{\partial^2 #1}{\partial #2^2}}

\begin{document}

\preprint{APS/123-QED}

\title{Stokes--anti-Stokes light scattering process: A photon-wave-function approach}% Force line breaks with \\
%\thanks{A footnote to the article title}%

\author{A. V. A. Guimar\~{a}es}
\email{afredovag@yahoo.com.br}
\affiliation{%
 Departamento de F\'{i}sica, ICEx, Universidade Federal de Minas Gerais.
Av. Antonio Carlos, 6627, Belo Horizonte, MG, 31270-901, Brazil}%
\author{Marcelo F. Santos}
\affiliation{Instituto de F\'{i}sica, Universidade Federal do Rio de Janeiro, CP68528, Rio de Janeiro, Rio de Janeiro 21941-972, Brazil}
\author{A. Jorio}%
\affiliation{%
 Departamento de F\'{i}sica, ICEx, Universidade Federal de Minas Gerais.
Av. Antonio Carlos, 6627, Belo Horizonte, MG, 31270-901, Brazil}%
\author{C. H. Monken}%
 \email{monken@fisica.ufmg.br}
\affiliation{%
 Departamento de F\'{i}sica, ICEx, Universidade Federal de Minas Gerais.
Av. Antonio Carlos, 6627, Belo Horizonte, MG, 31270-901, Brazil}%

\date{\today}

\begin{abstract}
The Photon wave function Formalism provides an alternative description of some quantum optical phenomena in a more intuitive way. We use this formalism to describe the process of correlated Stokes--anti-Stokes Raman scattering. In this process, two photons from a laser beam are inelastically scattered by a phonon created by the first photon (Stokes processes) and annihilated by the second photon  (anti-Stokes process), producing a Stokes--anti-Sokes (SaS) photon pair. We arrive at an expression for the two-photon wave function of the scattered SaS photon pair, which is in agreement with a number of experimental results.
\end{abstract}

%\keywords{Suggested keywords}
\maketitle

%\tableofcontents

\section{\label{intro}Introduction}
The concept of a Photon Wave Function (PWF) based on the Riemann-Silberstein vector, as introduced by Bialinycki-Birula and Sipe \cite{birula1, sipe, birula2} and extended by Smith and Raymer \cite{smith1}, provides an alternative description of some quantum optical phenomena in a more intuitive way \cite{smith2,keller1,keller2,birula3,correa,birula4,birula5}. When photons interact with matter, some caution must be exercised in the division of energy and momentum between the photon and the material medium. There is no unique prescription to do this division \cite{pablo2}  and different approaches to describe the propagation of photons in material media have been presented by Birula \cite{birula2}, Keller \cite{keller2} and Saldanha \cite{pablo2,pablo}. In this context, the PWF formalism seems to be the most convenient way to deal with near-field quantum electrodynamics, although in a fully covariant formulation in terms of the vector potential, instead of the Riemann-Silberstein vector \cite{keller2}.

In this work, we use the PWF formalism to describe the process of correlated Stokes and anti-Stokes Raman scattering, predicted by Klyshko in 1977 \cite{klyshko1}. In this process, two photons from a laser beam are inelastically scattered by a quantum of vibration (phonon) created by the first photon (Stokes process) and annihilated by the second photon  (anti-Stokes process), producing a Stokes--anti-Stokes (SaS) photon pair. The interest in SaS photon pairs has been increasing throughout the last ten years \cite{reim,reim2,lee,lee2,england,ado1,england2,kasperczyk,kasperczyk2,parra,pcp,anderson,velez,filo3,filo2}, carried by  recent developments of Quantum Optics and Quantum Information. Their interesting number- and time-correlation properties have been explored in solid-state quantum memories \cite{reim,reim2,england,england2,anderson}, preparation of macroscopic systems in entangled quantum states \cite{lee}, measurement of phonon coherence times \cite{lee2}, and preparation of a single quantum of vibration \cite{velez}. 

We start with a brief review of the PWF concept in the presence of matter, as introduced by Saldanha and Monken \cite{pablo}.  Next, we use the PWF formalism and the Green function method to describe the resonant SaS pair production, and finally, we present some conclusions and perspectives.

\section{\label{pwf}The Photon Wave Function}
The Bialynicki-Birula--Sipe PWF \cite{birula1,sipe} is a function of the position coordinates that completely describes the state of a photon.  However, an important difference between the PWF and the usual quantum-mechanical wave function of massive particles is that its modulus squared gives the photon energy density at a given point and not the probability density of finding the photon at that point.

We can define the PWF by means of the Riemann-Silberstein vector
\begin{equation}\label{rsv}
\begin{split}
\vec{\Psi}&=\vec{\Psi}_{+}+\vec{\Psi}_{-},\\
\vec{\Psi}_{\pm}&=\sqrt{\dfrac{\epsilon_0}{2}}\left(\vec{E}_{\pm}\pm ic\vec{B}_{\pm}\right),
\end{split}
\end{equation}
where the subscript $+\ (-)$ denotes positive (negative) helicity, $\vec{E}_{\pm}$ and $\vec{B}_{\pm}$ are real transverse (divergence-free) vector fields, $\vec{\Psi}_{\pm}$ are eingenstates of the the helicity operator $\hat{\sigma}$, satisfying $\hat{\sigma}\vec{\Psi}_{\pm}=\pm\vec{\Psi}_{\pm}$. It is important to notice that  if we expand $\vec{E}_{\pm}$ and $\vec{B}_{\pm}$ in terms of plane waves, 
\begin{align}
\vec{E}_{\pm}(\vec{r},t)&=i\sum_{\vec{k}}\Big[\vec{\mathcal{E}}_{\vec{k}\pm}e^{i(\vec{k}\cdot\vec{r}-\omega t)}-\vec{\mathcal{E}}^{*}_{\vec{k}\pm}e^{-i(\vec{k}\cdot\vec{r}-\omega t)}\Big],\\
\vec{B}_{\pm}(\vec{r},t)&=i\sum_{\vec{k}}\Big[\vec{\mathcal{B}}_{\vec{k}\pm}e^{i(\vec{k}\cdot\vec{r}-\omega t)}-\vec{\mathcal{B}}^{*}_{\vec{k}\pm}e^{-i(\vec{k}\cdot\vec{r}-\omega t)}\Big],
\end{align}
for each $\vec{k}$ one must have $c\vec{\mathcal{B}}_{\vec{k}\pm}=\uvec{k}\times\vec{\mathcal{E}}_{\vec{k}\pm}$. Since  $\uvec{k}\times\vec{\mathcal{E}}_{\vec{k}\pm}=\pm e^{-i\pi/2}\vec{\mathcal{E}}_{\vec{k}\pm}$,  $\vec{E}_{\pm}$ and $\pm c\vec{B}_{\pm}$ are pairs of Hilbert transforms. Therefore, $\vec{\Psi}$ is a complex analytic signal (positive-frequency) \cite{vakman,mandelwolf}.

$\vec{\Psi}$ obeys the following equations in free space:
\begin{align}\label{e2}
i\hbar\dfrac{\partial\vec{\Psi}}{\partial t}&=\hbar c\hat{\sigma}\nabla\times\vec{\Psi},\\
\label{e2b}
\nabla\cdot\vec{\Psi}&=0.
\end{align}

Saldanha and Monken derived a modified equation for $\vec{\Psi}$ in the presence of matter, which is justified by a particular division of the energy in electromagnetic and material parts \cite{pablo}. The result is an additional term in the right side of Eq.  \eqref{e2} that represents the response of the medium to the presence of a photon:
\begin{equation}\label{e5}
i\hbar\dfrac{\partial\vec{\Psi}}{\partial t}=\hbar c\hat{\sigma}\nabla\times\vec{\Psi}-i\dfrac{\vec{J}}{\sqrt{2\epsilon_0}},
\end{equation}
where $\vec{J}=\vec{J}_{\!f}+\vec{J}_{M}+\vec{J}_{P}$ is the current density induced by the photon, and
$\vec{J}_{\!f}$, $\vec{J}_{M}$ and $\vec{J}_{P}$ are due to free charges, magnetization and electric polarization, respectively. It is straightforward to show that Eqs. \eqref{e2b} and  \eqref{e5} are equivalent to
\begin{equation}
	\nabla^2\vec{\Psi}-\frac{1}{c^2}\dderivp{\vec{\Psi}}{t}=\sqrt{\frac{\mu_0}{2}}\Bigl(\frac{1}{c}\derivp{\vec{J}}{t}-i\hat{\sigma}\curl{\vec{J}}\Bigr).
\end{equation}

Here, we consider a nonmagnetic, transparent, nondispersive, homogeneous and isotropic medium in the absence of free charges.  In this case, $\vec{J}$ reduces to $\vec{J}_{P}=\partial\vec{P}/\partial{t}.$ The isotropy assumption is not really necessary and may be relaxed to include crystalline media, at the cost of a more complicated algebra.  We divide the response of the medium $\vec{P}=\chi_e\epsilon_0\vec{E}$ as $\vec{P}=\bar{\chi}_e\epsilon_0\vec{E}+\widetilde{\vec{P}}$, where $\bar{\chi}_{e}$ is a constant and  $\widetilde{\vec{P}}$ is due to any deviation of $\chi_{e}$ from  $\bar{\chi}_{e}$, which can be caused by nonlinearities or fluctuations. With this, we arrive at a wave equation for $\vec{\Psi}$
\begin{equation}
\label{psi6}
\nabla^2\vec{\Psi}-\frac{n^2}{c^2}\dderivp{\vec{\Psi}}{t}=\vec{f},
\end{equation}
 where $n=\sqrt{1+\bar{\chi}_e}$ is the refractive index and
\begin{equation}
\label{fonte}
\vec{f}=\mu_0\sqrt{\frac{\epsilon_0}{2}}\biggl(\dderivp{\widetilde{\vec{P}}}{t}-ic\sigma\curl{\derivp{\widetilde{\vec{P}}}{t}}\biggr).
\end{equation}
It is easy to show that $\vec{f}$ is also a complex analytic signal.

\section{\label{SAS}S\lowercase{a}S pair production}
To proceed with the calculation of the SaS pair production, we first have to derive an expression for $\widetilde{\vec{P}}$ and then apply second quantization to the fields.  Before that, some additional assumptions must be made about the system. First, we consider an active medium of volume $V\sub{S}$ composed of molecules having just one vibration mode with resonance frequency $\omega_{0}$. Second, in order to avoid internal reflections, we consider that $V\sub{S}$ is embedded in another linear medium of infinite volume, with the same refractive index $n$, whose molecular vibrational frequencies are far from $\omega_{0}$. This model can be extended to include   complex molecules with more than one resonant mode and crystal lattices supporting optical phonons.

\subsection{\label{pnlsection}The source term}
When a molecule interacts with a photon, its electronic cloud moves with respect to the nuclei, producing an induced electric dipole, which is, in first approximation, proportional to the electric field of the photon, $\vec{p}=\epsilon_0 \alpha\vec{E}$. However, instead of being constant, the polarizability $\alpha$ depends on the oscillation amplitude $Q$ of the nuclei with respect to their equilibrium position. Therefore, vibration of the nuclei will modify the molecular polarizability and cause a residual response of the molecules to the electric field.  We may write the polarizability $\alpha$ and the first-order residual response of the medium $\widetilde{\vec{P}}$, respectively, as
\begin{equation}
\label{pnl}
\begin{split}
\alpha&=\alpha_0+\deriv{\alpha}{Q}\Bigr|_{0}Q+...\\
\widetilde{\vec{P}}(\vec{r},t)&=N\epsilon_0 \alpha' Q(\vec{r},t)\vec{E}(\vec{r},t),
\end{split}
\end{equation}
where $\alpha'=d\alpha/dQ$ calculated at the equilibrium configuration ($Q=0$) and $N$ is the number of molecules per unit volume.  We will proceed in the scalar approximation, which is enough for our purposes, but $\alpha'$ is actually a tensor.

From now on we are going to treat each molecule as a quantum damped harmonic oscillator of resonance frequency $\omega_{0}$ and the medium in the continuum approximation as a molecular field \cite{klyshko2}. In order to take damping into account, we adopt a standard procedure and consider that the molecules interact with a  reservoir of harmonic oscillators with a broadband spectrum and a high density of frequency modes $\omega_{j}$. We assume that the reservoir oscillators do not interact with one another and the Markov approximation is valid, that is, the energy lost by the molecule to the reservoir never comes back and the molecule-reservoir interaction has no memory \cite{louisell,vankampen}. Of course, more sophisticated models are available, as for example in \cite{chines}. However, comparing our results with current experimental results, one can see that the Markov approximation  correctly describes the dynamics of the system under analysis.  In terms of the creation and annihilation operators $\operd{b}_{\vec{q}},\oper{b}_{\vec{q}}$ (molecular field) and $\operd{c}_{j\vec{q}},\oper{c}_{j\vec{q}}$ (reservoir) of vibration quanta in spatial modes $\exp(i\vec{q}\cdot\vec{r})$, the Hamiltonian of this composite system is given by 
\begin{equation}
\oper{H}=\oper{H}\subt{mol}+\oper{H}\subt{res}+\oper{H}\subt{int},	
\end{equation}
where
\begin{equation}
\oper{H}\subt{mol}=\half\hbar\omega_0\sum_{\vec{q}}\Big(\operd{b}_{\vec{q}}\oper{b}_{\vec{q}}+\oper{b}_{\vec{q}}\operd{b}_{\vec{q}}\Big)
\end{equation}
is the hamiltonian of the molecular oscillators, 
\begin{equation}
\oper{H}\subt{res}=\half\hbar\sum_{\vec{q}}\sum_{j}\Big[\omega_{ j }\Big(\oper{c}^{\dagger}_{j{\vec{q}}}\oper{c}_{j{\vec{q}}}+\oper{c}_{j{\vec{q}}}\oper{c}^{\dagger}_{j{\vec{q}}}\Big)\Big]
\end{equation}
is the hamiltonian of the set of oscillators composing the reservoir, and
\begin{equation}
\oper{H}\subt{int}=\hbar\sum_{\vec{q}}\sum_{j}\Big(\zeta_{j}^{*}\oper{c}^{\dagger }_{j{\vec{q}}}\oper{b}_{\vec{q}}+\zeta_{j}\oper{c}_{j{\vec{q}}}\operd{b}_{\vec{q}}\Big).
\end{equation}
is the interaction hamiltonian in the rotating-wave approximation. The coefficients  $\zeta_{j}$ account for the coupling strength between the molecules and the reservoir. 

In the Heisenberg picture, adopting the Weisskopf-Wigner approximation, the solution for $\oper{b}_{\vec{q}}(t)$ is given by \cite{louisell,vankampen}
\begin{equation}
\label{solb}
\begin{split}
\oper{b}_{\vec{q}}(t)=\oper{b}\sup{S}_{\vec{q}}e^{-i(\omegatil-i\gamma/2) t}+\oper{L}_{\vec{q}}(t),
\end{split}
\end{equation}
where $\oper{b}\sup{S}_{\vec{q}}=\oper{b}_{\vec{q}}(0)$ is independent of $t$ (Schr\"odinger operator), $\tilde{\omega}$ is the observed resonance frequency in the presence of damping, slightly deviated from $\omega_0$, $\gamma$ is the decay constant, and
\begin{equation}
\label{L}
\oper{L}_{\vec{q}}(t)=\sum_{j}\zeta_{j}\oper{c}_{j{\vec{q}}}\sup{S}\frac{e^{-i\omega_{j}t}-e^{-i(\omegatil-i\gamma/2)t}}{\omega_{j}-\omegatil+i\gamma/2}
\end{equation}
is a Langevin-type operator, with zero mean value on the reservoir, that ensures the commutation relation $[\oper{b}_{\vec{q}}(t),\operd{b}_{\vecp{q}}(t)]=\delta_{\vec{q},\vecp{q}}$ at any time.

The molecular vibration amplitude operator in terms of the creation and annihilation operators is given by
\begin{equation}
\label{qb}
\oper{Q}(\vec{r},t)=\sqrt{\frac{\hbar}{2M\omegatil}} \sum_{\vec{q}} \biggl[\oper{b}_{\vec{q}}(t)e^{i{\vec{q}}\cdot\vec{r}}+\operd{b}_{\vec{q}}(t)e^{-i{\vec{q}}\cdot\vec{r}}\biggr],
\end{equation} 
where $M$ is the total mass of the molecular oscillators.

We follow Ref. \cite{pablo} and write the electric field inside the medium as $\vec{E}=(\bar{\vec{\Psi}}+\bar{\vec{\Psi}}^{*})/\sqrt{2\epsilon}$, where
\begin{equation}
\label{psidressed}
\bar{\vec{\Psi}}(\vec{r},t)=\sqrt{\frac{\epsilon}{2}}\vec{E}(\vec{r},t)+i\hat{\sigma}\sqrt{\frac{\mu}{2}}\vec{B}(\vec{r},t)
\end{equation}
is the ``dressed photon'' wave function inside the medium.
Expanding $\bar{\vec{\Psi}}(\vec{r},t)$ in terms of energy eigenfunctions, we can convert it into a field operator (second quantization) as $\oper{\bar{\vec{\Psi}}}=\oper{\bar{\vec{\Psi}}}_{\!+}+\oper{\bar{\vec{\Psi}}}_{\!-}$, where
\begin{equation}
\oper{\bar{\vec{\Psi}}}_{\!h}(\vec{r},t)=	i\sum_{\vec{k}}\sqrt{\frac{\hbar\omega}{V\sub{Q}}}\, \oper{a}\sup{S}_{\vec{k}h}\uvec{e}_{\vec{k}h}e^{i(\vec{k}\cdot\vec{r}-\omega t)},
\end{equation}
$h=\pm$ denotes the helicity, $V\sub{Q}$ is the quantization volume, $\omega=u|\vec{k}|$, $u$ is the speed of light in the medium $c/n$, and $\oper{a}\sup{S}_{\vec{k}h}$ is the annihilation operator of a dressed photon in the corresponding plane-wave mode, satisfying $[\oper{a}\sup{S}_{\vec{k}h},\oper{a}^{{\scriptscriptstyle S}\dagger}_{\vecp{k} h'}]=\delta_{\vec{k},\vecp{k}}\delta_{h,h'}$. Notice that $u$ is a constant, since the medium is considered dispersionless. Using \eqref{psidressed}, we can write the electric field operator $\vec{\oper{E}}=\vec{\oper{E}}_{+}+\vec{\oper{E}}_{-}$, where
\begin{align}
\label{Efield}
\vec{\oper{E}}_{h}(\vec{r},t)&=i\sum_{\vec{k}}\sqrt{\frac{\hbar\omega}{2\epsilon V\sub{Q}}}\,\oper{a}\sup{S}_{\vec{k}h}\uvec{e}_{\vec{k}h}e^{i(\vec{k}\cdot\vec{r}-\omega t)}\nonumber\\
&+\mathrm{hermitian\ conjugate}.
\end{align}
From this point on, we drop the superscript $S$ of the time-independent operators.

Consistently with our definition of $\oper{\bar{\vec{\Psi}}}$ as a positive-frequency-only operator, we combine Eqs. \eqref{fonte}, \eqref{pnl}, \eqref{solb}--\eqref{Efield}, and define the positive-frequency source operator
\begin{align}
\vec{\oper{f}}(\vec{r},t)&=C\sum_{\vec{q}}\sum_{\vec{k}}\sqrt{\omega}\Big[\sum_{j}\frac{\zeta_{j}\,\oper{c}_{j{\vec{q}}}\,\vec{\oper{F}}_{\vec{k}+{\vec{q}},\omega+\omega_j}}{\omega_{j}-\omegatil+i\gamma/2}\nonumber\\
&\phantom{C\sum_{\vec{q}}\sum_{\vec{k}}\Big[}
+(\oper{b}_{\vec{q}}-\oper{v}_{\vec{q}})\vec{\oper{F}}_{\vec{k}+{\vec{q}},\Omega_a}\nonumber\\
&\phantom{C\sum_{\vec{q}}\sum_{\vec{k}}\Big[}
+\sum_{j}\frac{\zeta^{*}_{j}\,\operd{c}_{j{\vec{q}}}\,\vec{\oper{F}}_{\vec{k}-{\vec{q}},\omega-\omega_j}}{\omega_{j}-\omegatil-i\gamma/2}\nonumber\\
&\phantom{C\sum_{\vec{q}}\sum_{\vec{k}}\Big[}
+(\operd{b}_{\vec{q}}-\operd{v}_{\vec{q}})\vec{\oper{F}}_{\vec{k}-{\vec{q}},\Omega_s}\Big],
\end{align}
where $C=iN\alpha'\mu_0\epsilon_0\hbar/(2n\sqrt{2M\omegatil V\sub{Q}})$, $M$ is the total mass of the molecular oscillators, $\Omega_a=\omega+\omegatil-i\gamma/2$, $\Omega_s=\omega-\omegatil-i\gamma/2$,
\begin{displaymath}
\vec{\oper{F}}_{\vec{k},\Omega}=\Omega^2\,e^{i(\vec{k}\cdot\vec{r}-\Omega t)}\sum_{h}\oper{a}_{\vec{k}h}\Big(\uvec{e}_{\vec{k}h}+i\frac{c}{\Omega}\hat{\sigma}\vec{k}\times\uvec{e}_{\vec{k}h}\Big),
\end{displaymath}
and
\begin{displaymath}
\oper{v}_{\vec{q}}=\sum_{j}\frac{\zeta_{j}\oper{c}_{j{\vec{q}}}}{\omega_{j}-\omegatil+i\gamma/2}.
\end{displaymath}
It is easy to show that $[\oper{v}_{\vec{q}},\operd{v}_{\vecp{q}}]=\delta_{{\vec{q}},\vecp{q}}$ \cite{louisell}.
\subsection{\label{SaSgeneration} Photon scattering}

To calculate the SaS pair generation, we make use of the scattering theory by means of the dyadic Green function method, as we want to take all the polarization states into account. Assuming that one laser photon interacts with the medium,
\begin{equation}
\label{E0}
\oper{\bar{\vec{\Psi}}}(\vec{r},t)=\oper{\bar{\vec{\Psi}}}_{\!\ell}(\vec{r},t)+\oper{\bar{\vec{\Psi}}}_{\!\text{sc}}(\vec{r},t),
\end{equation}
where $\oper{\bar{\vec{\Psi}}}_{\!\ell}(\vec{r},t)$ corresponds to the laser photon. The scattered photon field operator is given by
\begin{equation}
\label{E1}
\oper{\bar{\vec{\Psi}}}_{\!\text{sc}}(\vec{r},t)=\int\limits_{V\sub{S}}\!d\vecp{r}\!\int\limits_{0}^{t}\!dt' \dyad{G}(\vec{r},t;\vecp{r},\tp)\cdot \vec{\oper{f}}(\vecp{r},\tp),
\end{equation}
where \cite{marx}
\begin{equation}
\label{gf}
\dyad{G}(\vec{r},t;\vecp{r},\tp)= \frac{1}{4\pi r}\delta[t'-(t_r+\uvec{r}{\cdot}\vecp{r}/u)](\dyad{I}-\uvec{r}\uvec{r})
\end{equation} 
is the appropriate retarded dyadic Green function in the far field approximation, $r=|\vec{r}|$, $\uvec{r}=\vec{r}/r$, $r'\ll r$, and $t_r=t-r/u$.
\begin{figure}
\begin{center}
\includegraphics[width=5.4cm]{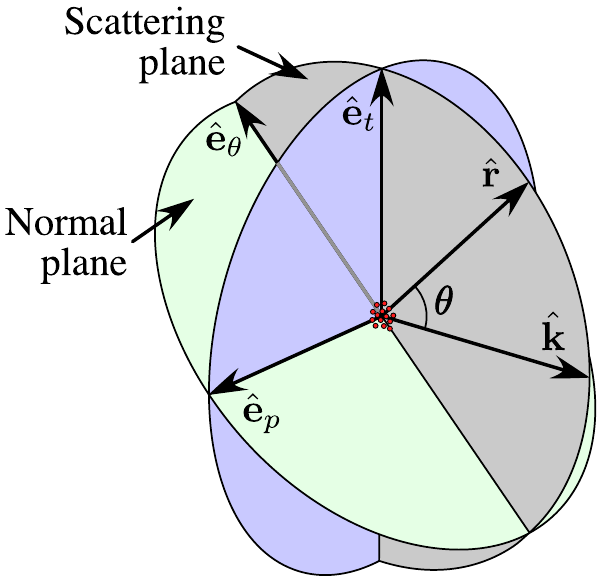}
\end{center}
\caption{\label{planes} Scattering geometry. $\uvec{k}$ indicates the laser photon propagation direction. $\uvec{r}$ indicates the observation direction. $\uvec{e}_{p}$, $\uvec{e}_{t}$ and $\uvec{e}_{\theta}$ are linear polarization vectors.}
\end{figure}
The dot products of the polarization vectors with the dyadic $(\dyad{I}-\uvec{r}\uvec{r})$ are
\begin{subequations}
\label{defek}
\begin{equation}
\uvec{e}_{\vec{k}h}\!\cdot(\dyad{I}-\uvec{r}\uvec{r})=\uvec{e}_{\vec{k}h}-(\uvec{e}_{\vec{k}h}\!\cdot \uvec{r})\uvec{r}=\vec{e}_{\vec{k}h}^{\perp},
\end{equation}
\begin{equation}
(\uvec{r}\times\uvec{e}_{\vec{k}h}){\cdot}(\dyad{I}-\uvec{r}\uvec{r})=\uvec{r}\times\vec{e}^{\perp}_{\vec{k}h},
\end{equation}
\end{subequations}
where $\vec{e}^{\perp}_{\vec{k}h}$ is the projection of $\uvec{e}_{\vec{k}h}$ on the plane normal to $\uvec{r}$. Notice that $\vec{e}^{\perp}_{\vec{k}h}$ is not normalized. The polarization vectors $\uvec{e}_{\vec{k}h}$ can always be written in terms of the basis vectors of linear polarization  $\uvec{e}_p,\uvec{e}_t$ (see Fig.\ref{planes})
\begin{equation}
\uvec{e}_{\vec{k}h}=\alpha_h\uvec{e}_p+\beta_h\uvec{e}_t,
\end{equation}
so that
\begin{equation}
\vec{e}^{\perp}_{\vec{k}h}=\alpha_h\uvec{e}_p+\beta_h\cos{\theta}\,\uvec{e}_\theta=A(\theta)\uvec{e}^{\perp}_{\vec{k}h},	
\end{equation}
where $\uvec{e}_p,\uvec{e}_\theta$ form a basis on the normal plane, $\theta$ is the angle between $\vec{r}$ and $\vec{k}$ and 
$A(\theta)=(\abs{\alpha_h}^2+\abs{\beta_h}^2\cos^2{\theta})^{1/2}$. 	$A(\theta)$ is independent of $h$, since $|\alpha_{+}|=|\alpha_{-}|$ and $|\beta_{+}|=|\beta_{-}|$.
Hence, $\uvec{e}^{\perp}_{\vec{k}h}$ is a polarization unit vector orthogonal to $\vec{r}$.

After integrating \eqref{E1}, supposing that $V\sub{S}$ is large enough to approximate $\int_{V\sub{S}}\exp[i(\vec{k}-\vecp{k})\cdot\vecp{r}]d\vecp{r}=V\sub{S}\delta_{\vec{k},\vecp{k}}$, the scattered field operator is
\begin{align}
\label{psi}
\oper{\bar{\vec{\Psi}}}_{\!\text{sc}}(\vec{r},t)&=D\sum_{\vec{k}}\sqrt{\omega}\bigg[\sum_{j}\frac{\zeta_{j}\oper{c}_{j\vec{k}^{+}_{j}-\vec{k}}\vec{\oper{F}}^{\perp}_{\vec{k},\omega^{+}_{j}}}{\omega_{j}-\omegatil+i\gamma/2}\nonumber\\
&\phantom{D\sum_{\vec{k}}\Bigl[}
+(\oper{b}_{\vec{k}_a-\vec{k}}-\oper{v}_{\vec{k}_a-\vec{k}})\vec{\oper{F}}^{\perp}_{\vec{k},\Omega_a}\nonumber\\
&\phantom{D\sum_{\vec{k}}\Bigl[}
+\sum_{j}\frac{\zeta^{*}_{j}\operd{c}_{j\vec{k}-\vec{k}^{-}_{j}}\vec{\oper{F}}^{\perp}_{\vec{k},\omega^{-}_{j}}}{\omega_{j}-\omegatil-i\gamma/2}\nonumber\\
&\phantom{D\sum_{\vec{k}}\Bigl[}
+(\operd{b}_{\vec{k}-\vec{k}_s}-\operd{v}_{\vec{k}-\vec{k}_s})\vec{\oper{F}}^{\perp}_{\vec{k},\Omega_s}\bigg],
\end{align}
where $D=CV\sub{S}/(2\pi)$, $\omega^{+}_{j}=\omega+\omega_{j}$, $\omega^{-}_{j}=\omega-\omega_{j}$, $\vec{k}^{+}_{j}=(\omega+\omega_j)\hat{\vec{r}}/u$, $\vec{k}^{-}_{j}=(\omega-\omega_j)\hat{\vec{r}}/u$, $\vec{k}_{a}=(\omega+\omegatil)\hat{\vec{r}}/u$, $\vec{k}_{s}=(\omega-\omegatil)\hat{\vec{r}}/u$, and
\begin{equation}
\label{Udef}
\begin{split}
\vec{\oper{F}}^\perp_{\vec{k},\Omega}(\vec{r},t)&=A(\theta)\frac{\Omega^2}{r}e^{-i\Omega t_r}\sum_{h}\oper{a}_{\vec{k}h}\,\uvec{e}^{\perp}_{\vec{k}h},
\end{split}
\end{equation}
where we have used the identity $i\hat{\sigma}(\uvec{r}\times\uvec{e}^{\perp})=\uvec{e}^{\perp}$.

Now, let us consider the initial state as $\oper{\rho}=\oper{\rho}\subt{em}\otimes\oper{\rho}\subt{m}$, where the electromagnetic part is  $\oper{\rho}\subt{em}=\proj{\psi\subt{em}}$ with $|\psi\subt{em}\rangle=|1_{\vec{k}_{\ell}\tau},1_{\vecp{k}_{\ell}\tau'}\rangle$, that is, a state with two laser photons, one in the mode $\vec{k}_{\ell}\tau$ and the other in the mode $\vecp{k}_{\ell}\tau'$, both with the same frequency $\omega_\ell$, not necessarily helicity eigenstates, and vacuum in all other electromagnetic modes. $\oper{\rho}\subt{med}=\oper{\rho}\subt{mol}\otimes\oper{\rho}\subt{res}$ is the vibrational state of the medium, where $\oper{\rho}\subt{mol}$ and $\oper{\rho}\subt{res}$ correspond to molecules and reservoir, respectively.  We are supposing that the process is parametric, that is, the state of the medium is not changed by the scattering process. In this case, the two-photon wave function is \cite{smith2}
\begin{equation}
\label{extract}
\vec{\Psi}\supt{(2)}(\vec{r}_1,\vec{r}_2,t)=\oper{S}\bra{0}\mathrm{Tr}\oper{\rho}\subt{m}\oper{\bar{\vec{\Psi}}}(\vec{r}_1,t)\oper{\bar{\vec{\Psi}}}(\vec{r}_2,t)\ket{\psi\subt{em}},
\end{equation}
where $\ket{0}$ is the electromagnetic vacuum state and $\oper{S}$ is the symmetrization operator. In thermal equilibrium, the states related to the medium are
\begin{equation}
\oper{\rho}\subt{mol}=(1-e^{-\tilde{\eta}})\sum_{\vec{q}}\sum_{n}e^{-\tilde{\eta}\,n_{\vec{q}}}	\proj{n_{\vec{q}}},
\end{equation}
where $\tilde{\eta}=\hbar\omegatil/(k\subt{B}T)$, $k\subt{B}$ is the Boltzmann constant, $n_{\vec{q}}$ is the number of phonons in mode $\vec{q}$ and
\begin{equation}
\oper{\rho}\subt{res}=\sum_{j}(1-e^{-\eta_{j}})\sum_{\vec{q}}\sum_{n_{j}}e^{-\eta_{j}n_{j\vec{q}}}	\proj{n_{j\vec{q}}},
\end{equation}
where $\eta_j=\hbar\omega_{j}/(k\subt{B}T)$.
The electromagnetic part of \eqref{extract} leads to
\begin{align}
&\bra{0}\!\sum_{h_1,h_2}\!\oper{a}_{\vec{k}_{1}h_{1}}\uvec{e}^{\perp}_{\vec{k}_{1}h_{1}}\oper{a}_{\vec{k}_{2}h_{2}}\uvec{e}^{\perp}_{\vec{k}_{2}h_{2}}\ket{\psi\subt{em}}=\nonumber\\
&
\delta_{\vec{k}_{1},\vec{k}_{\ell}}\delta_{\vec{k}_{2},\vecp{k}_{\ell}}\uvec{e}^{\perp}_{\vec{k}_{\ell}\tau}\uvec{e}^{\perp}_{\vecp{k}_{\ell}\tau'}+\delta_{\vec{k}_{1},\vecp{k}_{\ell}}\delta_{\vec{k}_{2},\vec{k}_{\ell}}\uvec{e}^{\perp}_{\vecp{k}_{\ell}\tau'}\uvec{e}^{\perp}_{\vec{k}_{\ell}\tau}.
\end{align}
We will assume that the average number of phonons at room temperature, given by $\mathcal{N}=(e^{\tilde{\eta}}-1)^{-1}$, is very low. In diamond, for instance, $\mathcal{N}\sim 10^{-3}$.  Then, we can approximate $\oper{\rho}\subt{mol}=\proj{0\subt{v}}$, where $\ket{0\subt{v}}$ is the vibrational vacuum state. We also assume that the system and the  reservoir are in thermal equilibrium, that is, the reservoir is also in the vacuum state. In this approximation, the material part of \eqref{extract} leads to the conditions
\begin{align}
\bra{0\subt{v}}\oper{c}_{j_1\vec{q}_1}\operd{c}_{j_2\vec{q}_2}\ket{0\subt{v}}&=\delta_{j_1,j_2}\delta_{\vec{q}_1,\vec{q}_2},\\
\bra{0\subt{v}}\oper{b}_{\vec{q}_1}\operd{b}_{\vec{q}_2}\ket{0\subt{v}} &=\delta_{\vec{q}_1,\vec{q}_2},\\
\bra{0\subt{v}}\oper{v}_{\vec{q}_1}\operd{v}_{\vec{q}_2}\ket{0\subt{v}} &=\delta_{\vec{q}_1,\vec{q}_2}.
\end{align}
 Therefore, the scattered part of \eqref{extract} is
\begin{align}
\label{psi2}
\vec{\Psi}\supt{(2)}_{\!\text{sc}}&=
D^2\omega_\ell\,\oper{S}'\bigg[\sum_{j}\frac{|\zeta_j|^2\vec{F}^{\perp}_{\vec{k}_\ell,\omega^{+}_{j}}\vec{F}^{\perp}_{\vecp{k}_\ell,\omega^{-}_{j}}}{(\omega_{j}-\omegatil)^2+\gamma^2/4}\delta_{\vec{k}^{+}_{j}+\vec{k}^{-}_{j},\vec{k}_\ell+\vecp{k}_\ell}\nonumber\\
&+2\,e^{-\gamma t}\,\vec{F}^{\perp}_{\vec{k}_\ell,\omega_{a}}\vec{F}^{\perp}_{\vecp{k}_\ell,\omega_{s}}\,\delta_{\vec{k}_{a}+\vec{k}_{s},\vec{k}_\ell+\vecp{k}_\ell}\bigg],
\end{align}
where the operator $\oper{S}'$ symmetrizes  $\vec{\Psi}_{\!\text{sc}}$ with respect to $\vec{r}_1,\vec{r}_2$ and $\vec{k}_{\ell}\tau,\vecp{k}_{\ell}\tau'$, $\omega^{+}_{j}=\omega_\ell+\omega_j$, $\omega^{-}_{j}=\omega_\ell-\omega_j$, $\omega_a=\omega_\ell+\omegatil$, $\omega_s=\omega_\ell-\omegatil$,
\begin{equation}
\vec{F}^{\perp}_{\vec{k}_\ell,\Omega}(\vec{r},t)=A(\theta)\frac{\Omega^2}{r}\,e^{-i\Omega t_{r}}\,\uvec{e}^{\perp}_{\vec{k}_{\ell}\tau},
\end{equation}
and $\theta$ is the angle between $\vec{r}$ and $\vec{k}_\ell$.

The fraction in the first term of \eqref{psi2} has a peak at $\omega_j=\omegatil$ and  width  $\gamma/2$, and a Kronecker delta that requires $\uvec{r}_1+\uvec{r}_2+(\uvec{r}_1-\uvec{r}_2)\omega_j/\omega_\ell=\uvec{k}_\ell+\uvecp{k}_\ell$. In general, $\gamma\ll\omegatil\ll \omega_\ell$, so that we can replace the Kronecker deltas by $\delta_{\uvec{r}_1+\uvec{r}_2,\uvec{k}_\ell+\uvecp{k}_\ell}$, $\omega_{j}^{+}$ by $\omega_{a}$ and $\omega_{j}^{-}$ by $\omega_{s}$. The first term can be further simplified if we replace $\sum_j\rightarrow\int \nu(\omega)\,d\omega$, where $\nu(\omega)$ is the number of frequency modes of the reservoir between $\omega$ and $\omega+d\omega$. Then,
\begin{align}
\label{int}
	&\sum_{j}\frac{|\zeta_j|^2\vec{F}^\perp_{\vec{k}_\ell,\omega^{+}_{j}}\vec{F}^\perp_{\vecp{k}_\ell,\omega^{-}_{j}}}{(\omega_{j}-\omegatil)^2+\gamma^2/4}\rightarrow A(\theta_1)A(\theta_2)\,\uvec{e}^{\perp}_{\vec{k}_{\ell} \tau}\uvec{e}^{\perp}_{\vecp{k}_\ell \tau'}\nonumber\\
&\times e^{i\omega_\ell t_{12}}|\zeta(\omegatil)|^2\nu(\omegatil)\,\frac{\omega_a^2\omega_s^2}{r_1r_2}\int\limits_{0}^{\infty}\frac{d\omega\,e^{i\omega(r_1-r_2)/u}}{(\omega-\omegatil)^2+\gamma^2/4},
\end{align}
where $t_{12}=(r_{1}+r_{2})/u$.
Extending the lower limit of the integral in \eqref{int} to $-\infty$ and taking into acount that $|\zeta(\omegatil)|^2\nu(\omegatil)=\gamma/(2\pi)$ \cite{louisell}, we can write
\begin{equation}
\sum_{j}\big[...\big]\rightarrow \,e^{-\gamma\delta t/2}\,\vec{F}^\perp_{\vec{k}_\ell,\omega_a}(\vec{r}_1,t)\vec{F}^\perp_{\vecp{k}_\ell,\omega_s}(\vec{r}_2,t),
\end{equation}
where $\delta t=|r_1-r_2|/u$.

Finally, in the stationary regime ($t\gg1/\gamma$), we get the two-photon scattered wave function
\begin{equation}
\label{Psifinal}
\begin{split}
\vec{\Psi}\supt{(2)}_{\!\text{sc}}(\vec{r}_1,\vec{r}_2,t)&=e^{-\gamma\delta t/2}\,\oper{S}'\,\big[\vec{\Psi}_{\!\omega_a}(\vec{r}_1,t)\vec{\Psi}_{\!\omega_s}(\vec{r}_2,t)\big]\\
&\times\delta_{\uvec{k}_\ell+\uvecp{k}_\ell,\uvec{r}_1+\uvec{r}_2},
\end{split}
\end{equation}
where $\vec{\Psi}_{\!\omega_a}(\vec{r}_1,t)$ and $\vec{\Psi}_{\!\omega_s}(\vec{r}_2,t)$ have the same expressions of far-field one-photon wave functions emitted by electric dipoles located at the origin, oscillating at frequencies $\omega_a$ and $\omega_s$, respectively \cite{novotny}, 
\begin{equation}
\vec{\Psi}\supt{(d)}_{\!\omega}(\vec{r},t)=\sqrt{2\epsilon_0}\Big[\frac{\mu_0}{4\pi}\,p\,A(\theta)\frac{\omega^2}{r}\,e^{-i\omega t_{r}}\uvec{e}^{\perp}_{\vec{k}_\ell \tau}\Big],
\end{equation}
with dipole moments given by
\begin{equation}
p=\frac{4\pi D\sqrt{\omega_\ell}}{\mu_0\sqrt{2\epsilon_0}}=\frac{\hbar NV\sub{S}\alpha'}{2n}\sqrt{\frac{\epsilon_0\omega_\ell}{M\omegatil V\sub{Q}}}.
\end{equation}

\section{\label{conclusion} Discussion and Conclusion}
Some facts about the result \eqref{Psifinal} are worth mentioning. First, it should be stressed that \eqref{Psifinal} is valid in the far-field approximation only. In the near-field domain, the situation is far more complex and requires another approach \cite{keller2}. Although \eqref{Psifinal} is written in terms of a product of two PWFs, the Kronecker delta imposes a strict angular correlation, meaning that the spatial coincidence detection profile of SaS photons follows the laser beam profile, as experimentally observed in \cite{filo3}. The polarization vectors of $\vec{\Psi}_{\!a}$ and $\vec{\Psi}_{\!s}$ are projections of the laser polarization vectors on the planes normal to $\vec{r}_1$ and to $\vec{r}_2$.  Then, if the laser photons have approximately the same polarization, as in a focused beam, the detection  probability of cross-polarized SaS photons is very small, as verified in \cite{filo2}. Finally, the decay term leads to a SaS pair detection probability decaying as $\exp{(-\gamma\delta t)}$ when a delay line is inserted in one of the detection paths. This decay is determined by the phonon lifetime $1/\gamma$, as verified in \cite{lee2,england,england2,anderson,velez,filo2}. This exponential decay is a consequence of the Lorentzian spectrum of the SaS photon pairs shown in Eqs. \eqref{psi2} and \eqref{int}, validating the assumed Weisskopf-Wigner approximation.

In conclusion, we provided an explanation for the SaS photon pair production phenomenon using the formalism of photon wave function and scattering theory. We derived an expression for the two-photon wave function of the scattered SaS photon pair, which is in agreement with a number of experimental results. Although we have considered the medium as nondispersive and isotropic, frequency dispersion and anisotropy can be included with an additional effort, using the appropriate Green functions in space-frequency domain and Raman susceptibility tensors. In a future publication, we will extend this formalism to explain the production of correlated SaS pairs in a continuum of frequencies, the photonic analogues of Cooper pairs, described in \cite{pcp}.  
\acknowledgments
This work was supported by  CNPq Projects 140887/2019-9, 307481/2013-1, 429165/2018-8, 302775/2018-8, 302872/2019-1, INCT-IQ 465469/2014-0, and FAPERJ project E-26/202.576/2019.  

%\bibliography{references}% Produces the bibliography via BibTeX.

%apsrev4-2.bst 2019-01-14 (MD) hand-edited version of apsrev4-1.bst
%Control: key (0)
%Control: author (8) initials jnrlst
%Control: editor formatted (1) identically to author
%Control: production of article title (0) allowed
%Control: page (0) single
%Control: year (1) truncated
%Control: production of eprint (0) enabled
%

\end{document}